\documentclass[lettersize,journal,draftclsnofoot,onecolumn,12pt]{IEEEtran}

\usepackage[caption=false]{}% Support for small, `sub' figures and tables

\usepackage{xcolor}

\usepackage{caption}
\usepackage{subcaption}

\usepackage{comment}
\usepackage{graphicx}
\usepackage{epstopdf}% To incorporate .png illustrations using PDFLaTeX, etc.
\usepackage{amsmath}
\DeclareMathOperator*{\argmin}{argmin}
\usepackage{algorithm}
\usepackage[noend]{algpseudocode}

\usepackage{amsfonts}
\usepackage{amssymb}
\usepackage{amsthm}
\usepackage[utf8]{inputenc}
\usepackage{enumerate}
\usepackage{cite}
\usepackage{graphicx}
\usepackage[utf8]{inputenc}
\usepackage{amssymb}
\usepackage{bbm}
\usepackage{mathtools}
\usepackage{mathdots}
\usepackage{comment}
\usepackage{textcomp}
\usepackage{url}
\usepackage{color,soul}
\usepackage{bbold}
\usepackage{caption}
\usepackage{subcaption}
\IEEEoverridecommandlockouts
\newtheorem*{remark}{Remark}
\newtheorem*{problem}{Problem}

\begin{document}

\title{Parsimonious System Identification from Fragmented Quantized Measurements}
\author{Omar M.Sleem and Constantino M. Lagoa\\ 
\small Dep. of Electrical Engineering, Pennsylvania State University, State College, PA 16801, USA. \\	
\small Email: \{\ oms46@psu.edu, cml18@psu.edu \}\
}
\maketitle

\begin{abstract}
Quantization is the process of mapping an input signal from an infinite continuous set to a countable set with a finite number of elements. It is a non-linear irreversible process, which makes the traditional methods of system identification no longer applicable. In this work, we propose a method for parsimonious linear time invariant system identification when only quantized observations, discerned from noisy data, are available. More formally, given a priori information on the system, represented by a compact set  containing the poles of the system, and quantized realizations, our algorithm aims at identifying the least order system that is compatible with the available information. The proposed approach takes also into account that the available data can be subject to fragmentation. Our proposed algorithm relies on an ADMM approach to solve a $\ell_{p},(0<p<1),$ quasi-norm objective problem. Numerical results highlight the performance of the proposed approach when compared to the $\ell_{1}$ minimization in terms of the sparsity of the induced solution. 
\end{abstract}


\section{Introduction}

\subsection{Motivation}
Quantization is the division of a quantity into a discrete number of small parts, often assumed to be integral multiple of a common quantity \cite{8,quan4}. A classical example of quantization by rounding off, for the application of estimating densities of histograms, was analyzed in \cite{quan2}. Since the processing of signals, i.e. speech and image, requires a digital environment, quantization plays an important role in bridging the analog and digital worlds \cite{quan3}. On one hand, quantization led to a new research area in control theory called network controlled systems (NCS) \cite{17}. NCS deals with the idea of controlling a process when the input and output signals are transmitted via a communication channel. On the other hand, it revealed the incompetence of the classical theory of system identification in considering quantized measurements \cite{17}. 

When a signal is subject to quantization, the quantization noise can no longer be modeled as a filtered white (zero mean and independent over time) noise and is signal dependent. Hence, in \cite{16,18} and references therein, the traditional theory of system identification was suggested to be modified to tackle the fact that the measurements are subject to quantization. 
Moreover, from \cite{21} (section 10.1), the classical identification procedures are not suitable for robust identification, when the signal is subject to quantization, because they identify a set of parameters of a fixed mathematical structure, where a fixed system order must be assumed.

Inspired by this, various works --which will be discussed in the next section in more detail-- explored the problem of system identification given quantized realizations. However, in this paper we aim to present a new approach to the problem of (Linear Time Invariant) LTI system identification from quantized outputs. This approach allows for the use of a priori information on the system and fragmented measurements of the output. In addition, our approach aims to recover the least order system that is compatible with the data by minimizing an $\ell_{p}$, $(0<p<1)$, quasi-norm objective.

The paper is organized as follows; in the remaining of the introduction, we provide a comprehensive discussion of the previous related work and our contribution. Section~\ref{Notations} introduces the notations that are used throughout the paper. In section \ref{system_description}, we thoroughly describe the system model used. The parsimonious system identification problem is formally provided in section \ref{problem_statment}. The proposed (Alternating Direction Method of Multipliers) ADMM algorithm based on $\ell_{p}$ quasi-norm approximation is described in section \ref{proposed_solution}. We validate our approach with an extensive suite of numerical simulations in section \ref{numerical_results}. Finally, the paper is concluded in section \ref{conclusion}.

\subsection{Related work}
The problem of simple representation of signals using quantization dates back to the 1940's and is one of the main threads of information theory \cite{quan1}. However, rigorous analysis did not begin until the 1980's. In \cite{quan_rel1,quan_ref2}, considering digital feedback control systems, the authors proposed a way in which one can specify system structures that alleviate the adverse effects of quantization. The works in \cite{quan_ref3,quan_ref4} demonstrate that quantization can induce a chaotic behavior in digital feedback systems. The results in \cite{quan_ref5,quan_ref6} are recognized as a quantum leap because the author was able to analyze the behavior of control systems in detail. The circumstances under which a discrete unstable LTI system can be stabilized, by choosing feedback control that depends on the quantized measurements, are studied. In \cite{quan_ref7}, the authors proposed a control design methodology, assuming a quantizer with variable sensitivity along with system state, that stabilizes LTI control systems with quantized measurements. In \cite{quan_ref8}, the coarsest quantizer that stabilizes a single input LTI system is shown to be a logarithmic one and can be obtained by solving a linear quadratic regulator problem. Abundant other works investigated the problem of the stabilization of NCS in different situations, e.g., \cite{quan_ref9,quan_ref10,quan_ref11}.

Despite that ample research activity in the stabilization and state estimation, quantization in system identification problems was still not properly pondered \cite{quan_ref12}. In \cite{15}, the authors studied the effect of quantization on I/O data used for system identification in a controlled plant whose parameters may change during the operation. They derived the optimal quantization scheme and showed that it is coarse near the origin of the signals and dense at a distance from it. This result is opposite to the case of stabilization in \cite{quan_ref8} and reveals duality between system identification and stabilization. Similar properties of the optimal quantizer were concluded in \cite{quan_ref13}, where the author considered a least square error objective function --for parameter estimation-- subject to a constraint on the number of subsections of the quantized signals or the expectation of the optimal code length for either high or low resolution.
In \cite{12}, the problem of system identification using uniformly quantized realizations was considered, where, the proposed formulation is a least square minimization of the difference equation errors over all time samples with the system parameters as optimization variables.
Regardless of the high accuracy in the estimation of the unknown information in the I/O data, the proposed method stills suffers the drawback of high computational complexity and noise neglection. The work in \cite{10} aimed to solve these drawbacks by exploiting statistical properties instead of deterministic treatment. In particular, an identification method for a linear system based on quantized measurements was derived. Using traditional equi-spaced quantizer, an instrumental level identification approach was proposed to enhance the estimation accuracy. The authors of \cite{9} took this approach a step further where a variation for the equi-spaced quantizer was considered. They showed that using a generalized noise shaping coder improves the accuracy of the estimates.

Another line of research includes the identification using a general class of quantized observations that allows the segmentation of the output range into a collection of subsets that may have unequal, fixed lengths or even design variables such as quantization design in communication systems and NCS \cite{14}. This serves in favor of understanding the potency of systems with limited sensor information, which in turns rapports the gap between resource limitations and identification complexity in sensor and communication networks. In particular, the work in \cite{13} considered the identification of a gain system by exploiting the information from multiple thresholds sensor and the convex combination of these thresholds. The results were extended to the case of a noisy communication channel through which the sensor output information is transmitted. The authors prove that their estimator is asymptotically efficient achieving the Cramer-Rao lower bound. Furthermore, the results were extended to a finite impulse response and transfer function models for periodic bounded input signals. In \cite{quan_ref14}, the authors focused on relationships between the identification space and time complexities. They showed that the asymptotic efficiency of empirical measure based algorithms yield to a tight bound on identification accuracy. This in turns aids to derive a separation principle of the complexities (time and space). The gained insights aim to provide a feasible approach for optimal utility of communication bandwidth resources in magnifying the identification accuracy. The role of dithering noise --adding artificial noise to the observed signal before quantization in order to mitigate the effects of quantization-- at the sensor was studied in \cite{11}. The authors asserted that tailored dithering noise can considerably simplify the derivation of optimal estimators in the expense of a decreased signal to noise ratio. 

\subsection{Contributions}
The different methods reviewed in the previous part aim to either stabilize the system, find an optimal quantization scheme or solve a system identification problem. In this work, we focus on the latter problem where, to the best of our knowledge, none of the proposed methods address the problem of identifying the system of least order that is compatible with collected information.

The problem of identifying systems using collected measurements can also include several other challenges including; 1) One can be faced with fragmented data due to the misplacement of sensors or external disturbances that can possibly make the collected data unreliable. 2) The ability to handle prior information on the system, e.g., constraints on the locations of the poles. 

In this paper, we aim to develop an algorithm that tackles the challenges mentioned above. More precisely, we consider a system in which the a priori information can be described by constraining the locations of the system's poles to be in a known compact set. Then, by exploiting ``simple representations'' of transfer functions, we develop an efficient algorithm that aims at finding the lowest order system that is compatible with fragmented quantized output measurements. This algorithm is based on an ADMM approach to the problem of $\ell_p$ quasi-norm optimization. To validate our analysis, we consider two different numerical examples; 1) System identification with randomly generated data set. 2) Identification with actual data collected from the motion of flexible robotic arm \cite{daisy_dataset}. The numerical results in both examples show that our method is competitive against the $\ell_{1}$ convex relaxation objective in terms of both the detected system order and accuracy of the recovered realizations. 

A preliminary version of part of this work was presented in \cite{CDC_paper}. This journal version includes a generalized formulation where a non continuous input data stream can be handled (input data is composed of independent chunks). Moreover, we do not assume a continuous measurement of the chunks' outputs from the quantizer, i,e,. output data is subject to fragmentation. Unlike \cite{CDC_paper}, we assume a generalized quantizer whose input is prone to noise and show that it plays an important role in the sparsity of the induced solution. We provide a new experiment to demonstrate the superior performance of our method in a more practical scenario.

\section{Notations} \label{Notations}
Unless otherwise specified, we denote scalars with non boldface letters, e.g., $x$, vectors with lowercase boldface letters, e.g., $\mathbf{x}$, with $i$-th entry as $x_{i}$, while matrices are in uppercase, e.g., $\mathbf{X}$, with $(i,j)$-th entry as $x_{i,j}$. $\mathbf{X}_{j,:}$ specifies the $j$-th row of the matrix $\mathbf{X}$.   $\mathbb{R}$ and $\mathbb{C}$ are the sets of real and complex numbers respectively. For a vector $\mathbf{x}$ and matrix $\mathbf{X}$, $|.|$ is an element-wise absolute value of the applied variable. However, for a set $\mathcal{X}$, $|.|$ operator stands for the cardinality of the set. We use $\preceq$ for element wise inequality of vectors. For any constant $c>0$, we define $\mathcal{I}_{c}\stackrel{\Delta}{=}[-c,c]$. For a positive integer $n$, we let $[n]\stackrel{\Delta}{=}\{1,\dots,n\}$. The $p$-th norm of a vector $\mathbf{x} \in \mathbb{R}^{n}$ is defined such that 
\begin{equation} \label{norm}
    \left\lVert \mathbf{x} \right\rVert_{p}\stackrel{\Delta}{=}(\sum_{i=1}^{n}|x_{i}|^{p})^{\frac{1}{p}}.
\end{equation}
It is important to note that when $0<p<1$, the expression in \eqref{norm} is a quasi-norm satisfying the same axioms of the norm except the triangular inequality making it a non-convex function.  For a complex number $x$, we use $\Bar{x}$ to denote the complex conjugate of that number. We let $\mathbf{1}$ be a vector of all entries equal to 1, $\mathbf{0}$ is a vector of zeros and $\mathbbm{1}_{\mathcal{X}}(.)$ be the indicator function to the set $\mathcal{X}$, i.e., it evaluates to zero if its argument belongs to the set $\mathcal{X}$ and is $+\infty$ otherwise. The compact set formed by the union of the interior and boundary of a unit circle, i.e. unit desk, on the complex domain centered around the origin is denoted by $\mathbb{D}$. Finally, for a matrix $\mathbf{X}$, we let $\text{vec}(\mathbf{X})$ be the vector formed by stacking its rows. 

\section{System description} \label{system_description}
\begin{figure}[t]
\begin{center}
\includegraphics[scale=.4]
{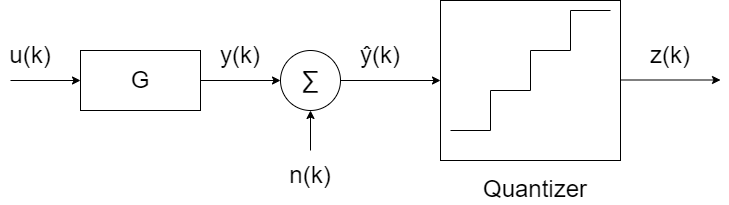}
\caption{System model.} \label{system_model}
\end{center}
\end{figure}

We consider the system shown in figure \ref{system_model}, where a discrete time input $u(k)$ on a finite time horizon is applied to a linear time invariant (LTI) system $G$. In control systems, the technology used to sense the process variable (output of the controlled process) often introduces noise, e.g., noise in electrical signals is due to interference from other electrical sources. We let a measurement noise $n(k) \in \mathcal{I}_{\epsilon}$ be added to the system output $y(k)$. The noisy output $\hat{y}(k)$ is then measured by the effect of a sensor that quantizes its input to discrete samples $\mathbf{z}(k)$. In the next part, we describe each component in figure \ref{system_model} thoroughly.

\subsection{LTI system $G$}
We consider a stable finite dimensional LTI system $G$ with poles that are contained in the compact set $\mathbb{D}$. The transfer function of the system, in the $z$-domain, can then represented as 
\begin{equation} \label{impulse_zdomain}
    H(z)=r+\sum_{q\in \mathbb{D}}\frac{a_{q}}{z-q},
\end{equation}
with $r\in \mathbb{R}$ and $a_{q} \in \mathbb{C}$ being the coefficient that is associated with pole $q$. For systems with repeated poles, an approximation by systems with transfer functions as in \eqref{impulse_zdomain} can be made with an arbitrary small precision level.

\subsection{Input data} \label{input_output}
The system $G$ models the relationship between the input $u(k)$ and the output signal $y(k)$. Besides boundness, we impose no constraints on the values of the samples of $u(k)$. The stability of the system $G$ ensures that the output $y(k)$ is bounded as long as $u(k)$ is bounded as well.

As mentioned before, and without loss of generality, we assume discrete time data with a sampling time of 1 unit. Moreover, we do not require continuous measurement of data. More precisely, input data is divided into multiple sets where continuous measurements are available. We refer to these sets as "chunks". 

\begin{figure}[t]
\begin{center}
\includegraphics[scale=.2]
{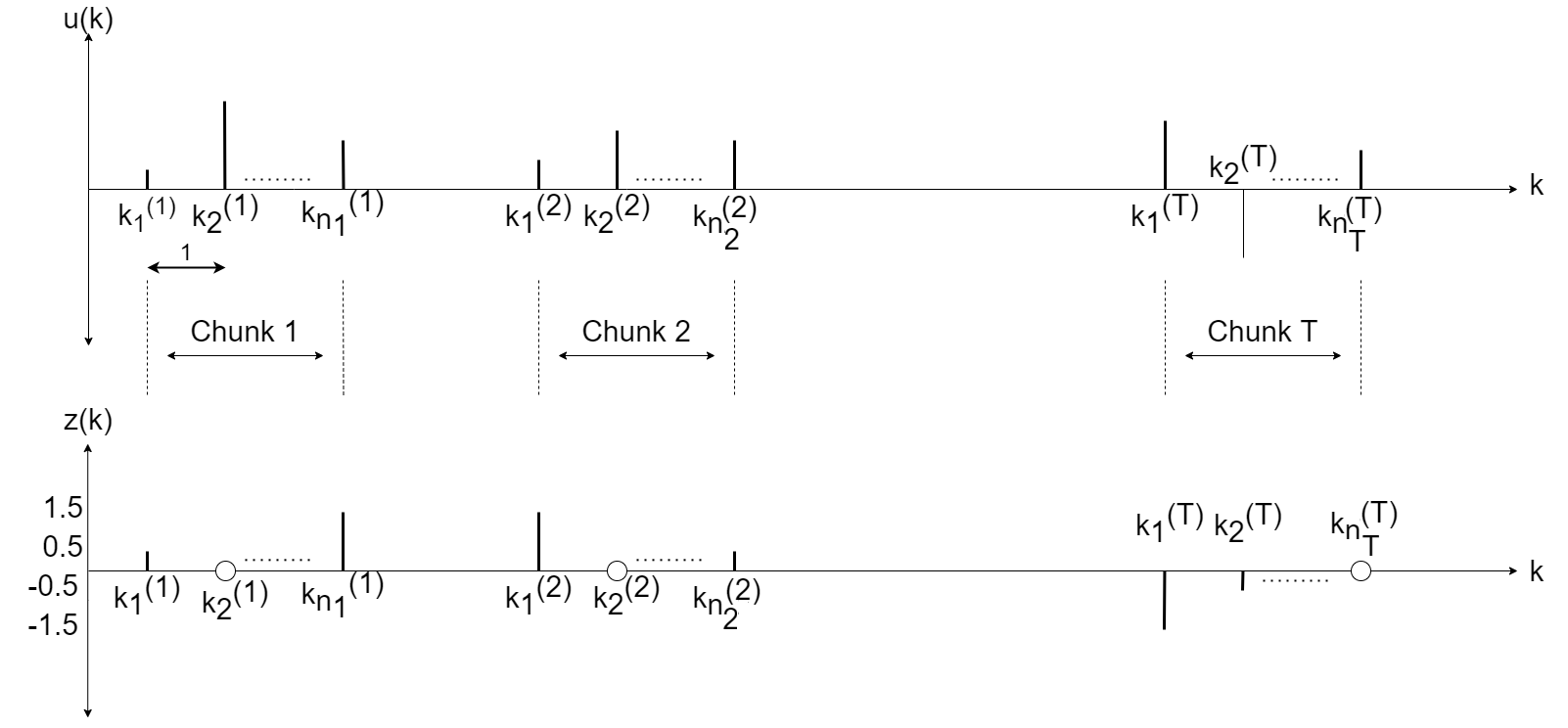}
\caption{Input/output data example. The circle indicates that the data is missing at that instance.} \label{input_output_data_ex}
\end{center}
\end{figure}

The upper part of figure \ref{input_output_data_ex} provides an input stream example, where $T$ different input data chunks, with size $n_{i}$ for chunk $i\in [T]$, are presented. For chunk $i$, the input sample at instance $k_{j}$ is denoted by $u(k_{j}^{(i)})$ where $j\in[n_{i}]$. Separations between different chunks as well as their sizes are arbitrary. 

\subsection{Quantizer and output data}
\begin{figure}[t]
\begin{center}
\includegraphics[scale=0.2]
{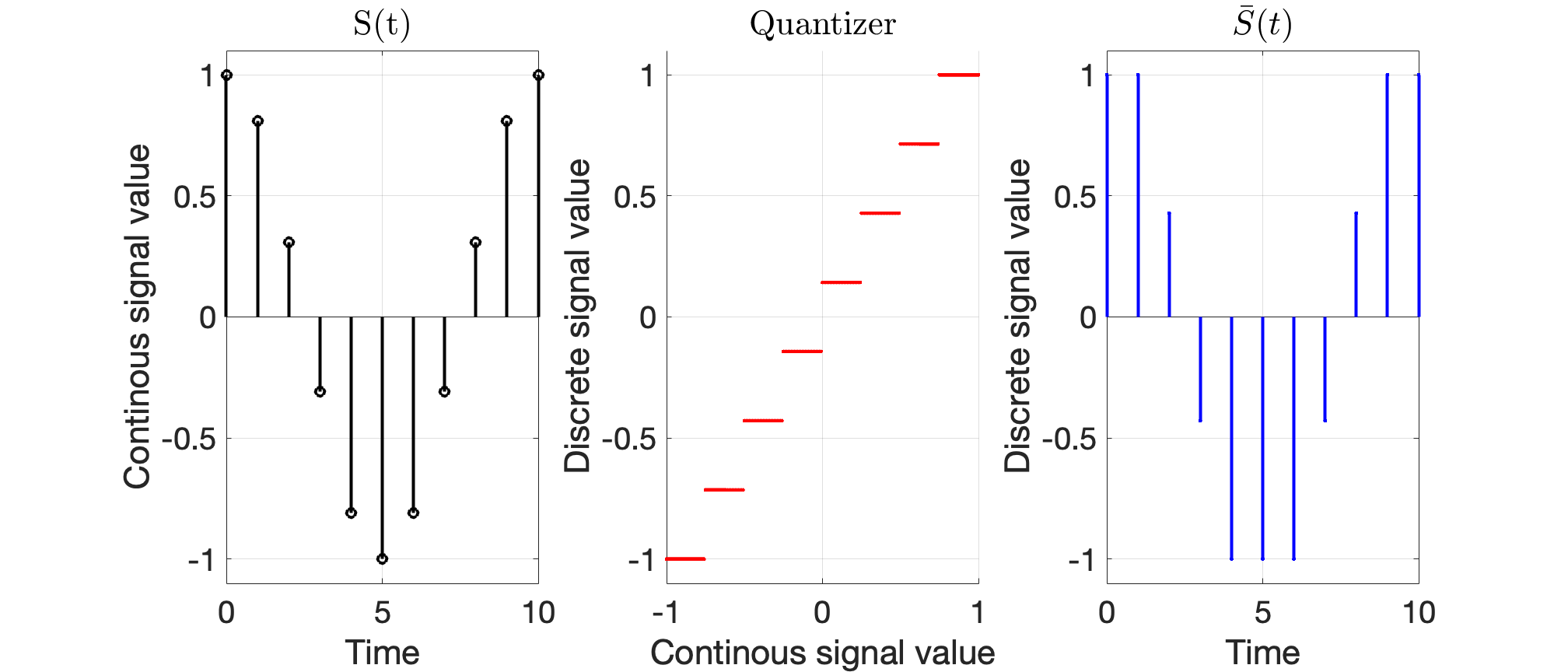}
\caption{Sensor operation example.} \label{sensor_operation}
\end{center}
\end{figure}

We assume a general quantizer, $Q$, that consists of the set of intervals $\mathcal{S}=\{\mathcal{S}_{i}, i\in \mathcal{I}\}$, with the index set $\mathcal{I}$ as ordinarily a collection of consecutive integers beginning with 1, together with a set of quantization levels $L=\{L_{i}, i\in \mathcal{I}\}$, so that the overall quantizer is defined by $Q(x)=L_{i}$ for $x\in \mathcal{S}_{i}$. The sets $\mathcal{S}_{i}$ partition the real line. That is, the cells are disjoint and exhaustive \cite{quan1}.
Without loss of generality, we assume a symmetric quantizer, where $L_{|\mathcal{I}|}=-L_{1}$ as the saturation level of the quantizer. Figure \ref{sensor_operation} provides an example for a uniform symmetric quantizer with $2^{m}$-levels, $m=3$, saturation value of 1 and a quantization step $\Delta=\frac{1}{2^{m-1}-0.5}=0.2857$. A cosine signal $S(t)$ is applied to the quantizer to produce the discrete signal $\Bar{S}(t)$. 

In addition, we do not assume that the all the output data stream is available within a chunk, i.e., the data is subject to fragmentation. This arises in cases when intermittent measurements are collected from the sensor or failure in communication occurs. 
%We also assume a digitalization hardware that associates every quantization level $L_{i}$ to a binary number starting from $(0\dots 0)$ for the first level to $(1\dots 1)$ for the last one. Throughout this paper, we let the quantizer output $\mathbf{z}(k) \in \{0,1\}^{m}$ for available data while $\mathbf{z}(k) \in \{-1\}^{m}$ for missing ones. 
The second part of Figure \ref{input_output_data_ex} provides an output example of a uniform $2^{m}$ levels sensor, with $m=2$ and a saturation level of 1.5, where the output for chunk $i\in [T]$ at instance $k_{j}$ is denoted by $z(k_{j}^{(i)})$ with $j\in[n_{i}]$.

As mentioned in the previous section, the input data stream (and correspondingly the output) chunks' separations are arbitrary and hence, we assume that the data from different chunks is independent. This, along with the time invariance assumption of the system, makes it reasonable to assume that the data chunks starting instances are the same, i.e., $k_{1}^{(i)}=1$ for all $i\in [T]$. For ease of notation, we drop the subscript and let $u^{(i)}(k)$ and $z^{(i)}(k)$ represent the input and output samples respectively of chunk $i\in [T]$ where $k\in [n_{i}]$.  

\section{Problem statement} \label{problem_statment}
Given input/output data $u^{(i)}(k)$ and $z^{(i)}(k)$, we aim to reconstruct the least order system that is compatible with the input output information and a priori assumptions on the system.
%Due to data fragmentation, the binary output $z(k)$ is only observed at discrete integer time instances defined over the set $\mathcal{K}\subseteq[\mathrm{N-1}]_{0}$.
More formally, the problem we aim to address can be stated as follows

\begin{problem}
Given 
\begin{itemize}
    \item Set $\mathbb{D}$ that contains the poles of the LTI system $G$.
    \item Input data chunks $u^{(i)}(k)$, $k\in [n_{i}]$,  $i\in [T]$, which are applied to the system $G$.
    \item A range $\mathcal{I}_{\epsilon}$ which includes the measurement noise $n^{(i)}(k)$, $k\in [n_{i}]$ and  $i\in [T]$. 
    \item Measurements of the fragmented sensor output realizations $z^{(i)}(k)$ for $k \in \mathcal{K}_{i} \subseteq [n_{i}]$.
\end{itemize}
find the most parsimonious system that is compatible with the a priori assumptions and a posteriori data mentioned above. 
\end{problem}
\begin{remark}
The formulation above assumes only the following a priori information, which are; 1)the system is stable and 2)the noise is bounded in $\mathcal{I}_{\epsilon}$. However, any other a priori information on the system $G$ that can be translated to constraints on the position of the poles (such as settling time), is compatible with the approach presented in this paper.
\end{remark}
\subsection{Parsimonious identification as a block sparsification problem}
From the definition of linear systems, the output at instance $k\in [n_{i}]$ within chunk $i\in [T]$, $y^{(i)}(k)$, can be decomposed as,
\begin{equation}
    y^{(i)}(k)=y_{\mathrm{zi}}^{(i)}(k)+y_{\mathrm{zs}}^{(i)}(k), 
\end{equation}
where, $y_{\mathrm{zi}}^{(i)}(k)$ is the zero input response at instance $k$ of chunk $i$, i.e., the response due to the initial conditions of the system before the input is applied, while $y_{\mathrm{zs}}^{(i)}(k)$ is the zero state response. From \cite{1}, the zero input response can be written as,
\begin{equation}
    y_{\mathrm{zi}}^{(i)}(k)=\sum_{q\in \mathbb{D}}b_{q}^{(i)}q^{k-1}, \quad \forall k \in [n_{i}], \quad \forall i \in [T],
\end{equation}
such that, similar to \eqref{impulse_zdomain}, $b_{q}^{(i)} \in \mathbb{C}$ is the coefficient that is associated to pole $q$ for chunk $i$. The zero state response is obtained by convolving the input sequence with the system's impulse response,
\begin{equation} \label{eq2}
    y_{\mathrm{zs}}^{(i)}(k)=\sum_{m=0}^{k}u^{(i)}(m)h(k-m), \quad \forall k \in [n_{i}], \quad \forall i \in [T],
\end{equation}
where $h(k)\stackrel{\Delta}{=}\mathcal{Z}_{z}^{-1}[H](k)$ is the system's impulse response and $\mathcal{Z}_{z}^{-1}[H](k)$ is the inverse $z$-transform of $H(z)$ with index $k$. By taking the inverse $z$-transform of \eqref{impulse_zdomain}, the impulse response can be easily found to be
%\begin{equation} \label{impulse_res}
%    h(k)=\delta(k)r+\sum_{n=1}^{\infty}\delta(k-n)\sum_{q\in \mathbb{D}}a_{q}q^{n-1},\\
%\end{equation} 
%\noindent where $\delta(k)$ is the dirac delta functional.
\begin{equation} \label{impulse_res}
    h(k)=\delta(k)r+\sum_{q\in \mathbb{D}}a_{q}q^{k-1} \text{step}(k-1),\\
\end{equation} 
\noindent where $\delta(k)$ is the dirac delta functional and $\text{step}(\cdot)$ is the step function defined as,
\[
\text{step}(k)=\begin{cases} 1 & \text{if } k\geq 0 \\  0 & \text{if } k<0  \end{cases}.
\]

Since system complexity and order are always related with the number of poles used to describe the system, we aim to reconstruct the system and the associated noise realization $n(k)$ for each sample, given only the quantized realizations $z(k)$, that can be depicted by the least number of poles. First, we let $\Upsilon:\mathbb{D}\rightarrow \mathbb{C}^{T+1}$ be the mapping from every pole $q$ to the corresponding coefficients $a_{q}$ and $b_{q}^{(i)}$, i.e., $\Upsilon(q)=\begin{bmatrix} a_{q}&b_{q}^{(1)}&\dots&b_{q}^{(T)}\end{bmatrix}^{\top}$. The problem mentioned earlier can then be formulated such that, for all $k\in\mathcal{K}_{i}$, $i\in[T]$ and $q\in \mathbb{D}$, we solve;
\begin{subequations}   \label{opt1}
\begin{align} 
\min_{a_{q},b_{q}^{(i)},r,n^{(i)}(k)}&\text{Cardinality}\{q\in\mathbb{D}:\Upsilon(q)\neq \mathbf{0}\}, \label{obj11} \\ 
\textrm{s.t.} \quad & y^{(i)}(k)=y_{\mathrm{zi}}^{(i)}(k)+y_{\mathrm{zs}}^{(i)}(k),\\ 
& y_{\mathrm{zi}}^{(i)}(k)=\sum_{q\in \mathbb{D}}b_{q}^{(i)}q^{k-1}, \label{con2}\\
& y_{\mathrm{zs}}^{(i)}(k)=\sum_{m=0}^{k}u^{(i)}(m)h(k-m), \\
& h(k)=\delta(k)r+\sum_{q\in \mathbb{D}}a_{q}q^{k-1} \text{step}(k-1), \label{con4}\\
& \hat{y}^{(i)}(k)=y^{(i)}(k)+n^{(i)}(k), \\
& z^{(i)}(k)\!=\!Q(\hat{y}^{(i)}(k)),  \\
& n^{(i)}(k) \in \mathcal{I}_{\epsilon}, \\
& a_{q}=\bar{a}_{\bar{q}}, \quad b_{q}^{(i)}=\bar{b}_{\bar{q}}^{(i)}. \label{con6}
\end{align} 
\end{subequations}
Constraint \eqref{con6} implies that the coefficients that are associated with complex conjugate poles have to be complex conjugate as well. 

\section{Proposed solution} \label{proposed_solution}
Theoretically, we aim to solve the problem in \eqref{opt1}. However, this is not feasible because the unit circle contains an infinite number of poles which makes the computational complexity of the problem intractable. We aim to implement an approximation of the above problem which is based on using a grid of the unit circle of size $n$. The denser the grid, the more accurate the approximation is to the original problem. However, a trade-off could exist as it increases the problem's computational complexity. First, we define the vector $\mathbf{q}^{\top}\!\!=\![q_{1},\dots q_{n}]$, which is composed of complex conjugates and real poles resulted from the gridding effect, the vector of the associated zero state coefficients $\mathbf{a}^{\top}\!\!=\!\![a_{q_{1}},\dots a_{q_{n}}]$ and the matrix of zero input coefficients $\mathbf{B}\in\mathbb{C}^{T\times n}$, where $\mathbf{B}_{i,:}\!\!=\![b_{q_{1}}^{(i)},\dots b_{q_{n}}^{(i)}]$. 
We also let $\mathbf{n}_{i}\in \mathbb{R}^{|\mathcal{K}_{i}|}$ be the vector of noise realizations $n^{(i)}(k)$ for $k\in \mathcal{K}_{i}$ with chunk $i\in [T]$. 
%As mentioned before, without loss of generality, we assume that the sizes of the data chunks are equal and the number of missing samples per chunks are also equal while their instances are different. Hence, we let $n_{i}=n_{t}=\eta$ for all $i,t\in [T]$ and $|\mathcal{K}_{i}|=|\mathcal{K}_{t}|=\kappa$ for all $i,t \in [\eta]$. We let the matrix $\mathbf{N}\in \mathbb{R}^{T\times \eta}$ be the matrix of all possible noise realizations on the available time domain. 

Second, we aim to equalize the energy contribution of all the poles and hence, we let the scaling factor $\boldsymbol\alpha \in \mathbb{R}^{n}$ be defined as, 
\begin{equation} \label{alpha}
    \alpha_{m}=\frac{1-|q_{m}|^{2}}{1-|q_{m}|^{2N+2}} \quad \forall m \in [n].
\end{equation}
The scaling factor $\boldsymbol\alpha$ aims to make the Hankel matrix formed by the system's impulse response has a nuclear norm equal to 1. For more information on $\boldsymbol\alpha$ and its proper choice, the interested reader is referred to \cite{19}. A good approximation for the problem in \eqref{opt1} can then be defined such that, for all $k\in\mathcal{K}_{i}$, $i\in[T]$ and $j\in [n]$, we aim to solve;
\begin{subequations}   \label{opt2}
\begin{align}  
\min_{\mathbf{a},\mathbf{B},r,\mathbf{n}_{i},\mathbf{d}}&\left\lVert \mathbf{d}\right\rVert_{0},\label{obj}\\ %\stackrel{\Delta}{=}\sum_{i=1}^{n}d_{i}^{p}, \label{obj}\\ 
\textrm{s.t.} \quad & %y(k)\!=\!(\boldsymbol\alpha\odot\mathbf{b})^{\top}\mathbf{q}^{\{k-1\}}\!+\!\sum_{j=0}^{k}u(j)h(k-j), \label{con1_2} \\
y^{(i)}(k)=y_{\mathrm{zi}}^{(i)}(k)+y_{\mathrm{zs}}^{(i)}(k), \label{con1_2}\\ 
& y_{\mathrm{zi}}^{(i)}(k)=\sum_{j\in [n]}\alpha_{j}b_{q_{j}}^{(i)}q_{j}^{k-1}, \label{con2_2}\\
& y_{\mathrm{zs}}^{(i)}(k)=\sum_{m=0}^{k}u^{(i)}(m)h(k-m), \label{con3_2}\\
& h(k)=\delta(k)r+\sum_{q\in \mathbb{D}}a_{q}q^{k-1} \text{step}(k-1), \label{con4_2}\\
& \hat{y}^{(i)}(k)=y^{(i)}(k)+n^{(i)}(k), \label{con5_2}\\
& z^{(i)}(k)\!=\!Q(\hat{y}^{(i)}(k)),  \\
& n^{(i)}(k) \in \mathcal{I}_{\epsilon}, \\
& a_{q_{j}}=\bar{a}_{\bar{q}_{j}}, \quad b_{q_{j}}=\bar{b}_{\bar{q}_{j}},  \label{con6_2} \\
& |\mathbf{a}|\preceq \mathbf{d}, \quad |\mathbf{B}_{i,:}|\preceq \mathbf{d}. \label{con7_2}
\end{align} 
\end{subequations}
The auxiliary variable $\mathbf{d}\in \mathbb{R}_{+}^{n}$ ensures block sparsity of the zero state and zero input coefficients, i.e., $\mathbf{a}$ and $\mathbf{B}$. A proper choice of the vector $\boldsymbol\alpha$, defined in \eqref{alpha}, and the use of \eqref{obj} and \eqref{con7_2} allow the identification of the system with the least number of poles, i.e., least order system. However, the $\ell_{0}$ pseudo-norm is an NP hard problem and hence, using notions of sparsity \cite{2}, the objective function is relaxed using the $\ell_{p}(0<p<1)$ quasi-norm, i.e., $\left\lVert \textbf{d}\right\rVert_{0}$ in \eqref{obj} is replaced with $\left\lVert \textbf{d}\right\rVert_{p}^{p}$ defined as in \eqref{norm}.

For notation simplicity, we define the vector $\mathbf{w} \in \mathbb{C}^{1+n(T+1)+\sum_{i\in [T]}|\mathcal{K}_{i}|}$, which is the concatenation of the variables $r$, $\mathbf{a}, \text{vec}(\mathbf{B})$ and $\mathbf{n}_{i}$ for $i\in [T]$. Let the set $\mathcal{D}\subseteq \mathbb{C}^{1+n(T+1)+\sum_{i\in [T]}|\mathcal{K}_{i}|}\times \mathbb{R}_{+}^{n}$ as the set of doubles $(\mathbf{w},\mathbf{d})$ where constraints \eqref{con1_2} to \eqref{con7_2} are satisfied. Hence, the problem in \eqref{opt2}, after the objective function relaxation, will have the compact representation in the form;
\begin{subequations}   \label{opt3}
\begin{align}  
\min_{\mathbf{w},\mathbf{d}} \quad & \left\lVert \textbf{d}\right\rVert_{p}^{p}, \label{obj2}\\ 
\textrm{s.t.} \quad &\mathbf{w},\mathbf{d} \in \mathcal{D}.
\end{align} 
\end{subequations}

As discussed, we aim to recover the lowest order system and hence, we consider the case when $0<p<1$, which lead to a non-convex objective in \eqref{opt3}. In our anaylysis, we consider an ADMM approach that utilizes the structure of the problem in order to divide the optimization over the variables via iteratively solving simpler sub-problems. 
Starting with the epi-graph form of \eqref{opt3} through introducing the auxiliary variable $\mathbf{t}\in \mathbb{R}^{n}$, where, 
\begin{align}   \label{opt3_d}
\min_{\mathbf{w},\mathbf{d},\mathbf{t}} \quad & \mathbf{1}^{\top} \mathbf{t}, \\ 
\textrm{s.t.} \quad &t_{i} \geq |d_{i}|^{p}, \quad i \in [n], \nonumber \\
&\mathbf{w},\mathbf{d} \in \mathcal{D} \nonumber.
\end{align} 
Let the non-convex set $\mathcal{X} \subset \mathbb{R}^{2}$ be the epigraph of the scalar function $|d|^{p}$, i.e., $\mathcal{X}=\{(d,t) \in \mathbb{R}^2: t \geq |d|^{p}\}$. Then, \eqref{opt3_d} can be cast as
\begin{align}   \label{opt4}
\min_{\mathbf{w},\mathbf{d},\mathbf{t}} \quad & \sum_{i \in [n]} \mathbbm{1}_{\mathcal{X}}(d_{i},t_{i}) + \mathbf{1}^{\top}\mathbf{t}, \\ 
\textrm{s.t.} \quad &\mathbf{w},\mathbf{d} \in \mathcal{D}. \nonumber 
\end{align} 
In order to write \eqref{opt4} in an ADMM form, we introduce the variables $\mathbf{s} \in \mathbb{C}^{1+n(T+1)+\sum_{i\in [T]}|\mathcal{K}_{i}|}$, $\mathbf{f}$ and $\mathbf{z} \in \mathbb{R}^{n}$, and hence, an equivalent ADMM formulation can be then given by:
\begin{align}  \label{lp_ADMM} 
\min_{\mathbf{w},\mathbf{d},\mathbf{t},\mathbf{s},\mathbf{f},\mathbf{z}} \quad & \sum_{i \in [n]} \mathbbm{1}_{\mathcal{X}}(d_{i},t_{i}) + g_{\mathcal{D}}(\mathbf{s},\mathbf{f})+\mathbf{1}^{\top}\mathbf{z}, \\ 
\textrm{s.t.} \quad & \mathbf{w}=\mathbf{s}: \quad \boldsymbol\lambda_{1}, \nonumber \\
& \mathbf{d}=\mathbf{f}: \quad \boldsymbol\lambda_{2}, \nonumber \\
&\mathbf{t}=\mathbf{z}: \quad \boldsymbol \theta. \nonumber
\end{align} 
The dual variables associated with the constraints $\mathbf{w}=\mathbf{s}$, $\mathbf{d}=\mathbf{f}$ and $\mathbf{t}=\mathbf{z}$ are $\boldsymbol\lambda_{1}$, $\boldsymbol\lambda_{2}$ and $\boldsymbol\theta$, respectively. Hence, the Lagrangian function corresponding to (\ref{lp_ADMM}) augmented with a quadratic penalty on the violation of the equality constraints with penalty parameter $\rho > 0$, is given by:
\begin{align}\label{Lagrangian}
&\mathcal{L}_{\rho}(\mathbf{d},\mathbf{t},\mathbf{s},\mathbf{f},\mathbf{w},\mathbf{z},\boldsymbol\lambda_{1},\boldsymbol\lambda_{2},\boldsymbol\theta)=
\sum_{i \in [n]} \!\mathbbm{1}_{\mathcal{X}}(d_{i},t_{i}) + g_{\mathcal{D}}(\mathbf{s},\mathbf{f})+\notag\\ 
&\mathbf{1}^{\top}\mathbf{z}+\boldsymbol\lambda_{1}^{\top}\!(\mathbf{w}-\mathbf{s})+\boldsymbol\lambda_{2}^{\top}\!(\mathbf{d}-\mathbf{f}) 
+ \boldsymbol\theta^{\top}\!(\mathbf{t}-\mathbf{z}) + \frac{\rho}{2} (\|\mathbf{w}-\mathbf{s}\|_{2}^{2} \notag \\ &+\|\mathbf{d}-\mathbf{f}\|_{2}^{2}+\|\mathbf{t}-\mathbf{z}\|_{2}^{2} ). 
\end{align}

Considering the three block variables $\mathbf{Q}_{1}=(\mathbf{d},\mathbf{t})$, $\mathbf{Q}_{2}=(\mathbf{s},\mathbf{f})$ and $\mathbf{Q}_{3}=(\mathbf{w},\mathbf{z})$, ADMM \cite{3} consists of the following iterations, where $l$ is the iteration number:
\begin{comment}
\begin{eqnarray}
(\mathbf{d},\mathbf{t})^{k+1} &\!=&\!\underset{\mathbf{d},\mathbf{t}}{\mathrm{argmin}}~
\mathcal{L}_{\rho}(\mathbf{w}^{k},\mathbf{d},\mathbf{t},\mathbf{s}^{k},\mathbf{f}^{k},\mathbf{z}^{k},\boldsymbol\lambda_{1}^{k},\boldsymbol\lambda_{2}^{k},\boldsymbol\theta^{k}) \label{up1}\\
(\mathbf{s},\mathbf{f})^{k+1}&\! = &\! \underset{\mathbf{s},\mathbf{f}}{\mathrm{argmin}}~
 \mathcal{L}_{\rho}(\mathbf{w}^{k},\mathbf{d}^{k+1},\mathbf{t}^{k+1},\mathbf{s},\mathbf{f},\mathbf{z}^{k},\boldsymbol\lambda_{1}^{k},\boldsymbol\lambda_{2}^{k},\boldsymbol\theta^{k}) \label{up3}\\
 (\mathbf{w},\mathbf{z})^{k+1}&\! = &\! \underset{\mathbf{w},\mathbf{z}}{\mathrm{argmin}}~
 \mathcal{L}_{\rho}(\mathbf{w},\mathbf{d}^{k+1},\mathbf{t}^{k+1},\mathbf{s}^{k+1},\mathbf{f}^{k+1},\mathbf{z},\boldsymbol\lambda_{1}^{k},\boldsymbol\lambda_{2}^{k},\boldsymbol\theta^{k}) \label{up2}\\
 \boldsymbol\lambda_{1}^{k+1} &\! = &\! \boldsymbol\lambda_{1}^{k}+\rho(\mathbf{w}^{k+1}-\mathbf{s}^{k+1})  \label{ADMM_lambda}, \\
 \boldsymbol\lambda_{2}^{k+1} &\! = &\! \boldsymbol\lambda_{2}^{k}+\rho(\mathbf{d}^{k+1}-\mathbf{f}^{k+1})  , \\
 \boldsymbol\theta_{1}^{k+1} &\! = &\! \boldsymbol\theta_{1}^{k}+\rho(\mathbf{t}^{k+1}-\mathbf{z}^{k+1}).  \label{ADMM_theta}
\end{eqnarray}
\end{comment}
\begin{eqnarray}
\mathbf{Q}_{1}^{\{l+1\}}&\!\!\!\!\!\!=&\!\!\!\!\!\underset{\mathbf{d},\mathbf{t}}{\mathrm{argmin}}~\!
\mathcal{L}_{\rho}(\mathbf{Q}_{1},\mathbf{Q}_{2}^{\{l\}},\mathbf{Q}_{3}^{\{l\}}\!,\!\boldsymbol\lambda_{1}^{\{l\}}\!,\!\boldsymbol\lambda_{2}^{\{l\}}\!,\!\boldsymbol\theta^{\{l\}}), \label{up1}\\
\mathbf{Q}_{2}^{\{l+1\}}&\!\!\!\!\!\!=&\!\!\!\!\!\underset{\mathbf{s},\mathbf{f}}{\mathrm{argmin}}~\!
 \mathcal{L}_{\rho}(\mathbf{Q}_{1}^{\{l+1\}}\!,\!\mathbf{Q}_{2},\!\mathbf{Q}_{3}^{\{l\}}\!,\!\boldsymbol\lambda_{1}^{\{l\}}\!,\!\boldsymbol\lambda_{2}^{\{l\}}\!,\!\boldsymbol\theta^{\{l\}}), \label{up3}\\
 \mathbf{Q}_{3}^{\{l+1\}}&\!\!\!\!\!\!=&\!\!\!\!\!\underset{\mathbf{w},\mathbf{z}}{\mathrm{argmin}}~\!
 \mathcal{L}_{\rho}(\mathbf{Q}_{1}^{\{l+1\}}\!,\!\!\mathbf{Q}_{2}^{\{l+1\}}\!,\!\!\mathbf{Q}_{3},\!\boldsymbol\lambda_{1}^{\{l\}}\!,\!\boldsymbol\lambda_{2}^{\{l\}}\!,\!\boldsymbol\theta^{\{l\}}), \label{up2} \\
 \boldsymbol\lambda_{1}^{\{l+1\}} &\!\!\!\!\!\!=&\!\!\!\!\!\boldsymbol\lambda_{1}^{\{l\}}+\rho(\mathbf{w}^{\{l+1\}}-\mathbf{s}^{\{l+1\}})  \label{ADMM_lambda}, \\
 \boldsymbol\lambda_{2}^{\{l+1\}} &\!\!\!\!\!\!=&\!\!\!\!\!\boldsymbol\lambda_{2}^{\{l\}}+\rho(\mathbf{d}^{\{l+1\}}-\mathbf{f}^{\{l+1\}})  , \\
 \boldsymbol\theta_{1}^{\{l+1\}} &\!\!\!\!\!\!=&\!\!\!\!\!\boldsymbol\theta_{1}^{\{l\}}+\rho(\mathbf{t}^{\{l+1\}}-\mathbf{z}^{\{l+1\}}).  \label{ADMM_theta}
\end{eqnarray}
\subsection{($\mathbf{d}$, $\mathbf{t})$ update}
From the expression of the augmented Lagrangian in \eqref{Lagrangian} and by completing the square, the update of $\mathbf{d}$ and $\mathbf{t}$ in \eqref{up1} can be found by solving the following optimization,
\begin{equation}  \label{d,t_up}
\begin{aligned}
\min_{\mathbf{d},\mathbf{t}} \quad & \|\mathbf{d}-(\mathbf{f}^{\{l\}}-\frac{\boldsymbol\lambda_{2}^{\{l\}}}{\rho})\|_{2}^{2}+\|\mathbf{t}-(\mathbf{z}^{\{l\}}-\frac{\boldsymbol\theta^{\{l\}}}{\rho})\|_{2}^{2}, \\
\textrm{s.t.} \quad & (d_{i},t_{i}) \in \mathcal{X} \quad \forall i \in [n].\\
\end{aligned}
\end{equation}

It can be realized that the problem in \eqref{d,t_up} enjoys a separable structure and hence is amenable to decentralization. However, it is a non-convex problem due to the nature of the set $\mathcal{X}$. In \cite{4}, the authors considered a similar problem and it was shown that the element-wise optimization of \eqref{d,t_up} boils down to finding the roots, $a_{i}^{*}$, of the scalar $2v$ polynomial;
\begin{equation}\label{polynomial}
a_{i}^{2v}+\frac{u}{v}\left(a_{i}^{2u}-\tilde{t}_{i}a_{i}^{u}\right)-\tilde{x}_{i}a_{i}^{v},
\end{equation}
where $\tilde{x}_{i}=f_{i}^{\{l\}}-\frac{\lambda_{i,2}^{\{l\}}}{\rho}$, $\tilde{t}_{i}=z_{i}^{\{l\}}-\frac{\theta_{i}^{\{l\}}}{\rho}$ and $u,v \in \mathbb{Z}_{+}$ such that $p=u/v$. They showed that, in proposition 1, the entry-wise solution of \eqref{d,t_up} is given by $(d_{i}^{*},t_{i}^{*})=(a_{i}^{*^{v}},a_{i}^{*^{u}})$ for all $i \in [n]$. 
\subsection{($\mathbf{s}$, $\mathbf{f})$ update}
By fixing all the remaining variables, the ($\mathbf{s}$, $\mathbf{f})$ update in \eqref{up3} can be easily shown to be the solution of the following optimization problem;
\begin{equation}  \label{s,f_up}
\begin{aligned}
\min_{\mathbf{s},\mathbf{f}} \quad & \|\mathbf{s}\!-\!(\mathbf{w}^{\{l\}}\!+\!\frac{\boldsymbol\lambda_{1}^{\{l\}}}{\rho})\|_{2}^{2}\!+\!\|\mathbf{f}\!-\!(\mathbf{d}^{\{l+1\}}+\frac{\boldsymbol\lambda_{2}^{\{l\}}}{\rho})\|_{2}^{2}, \\
\textrm{s.t.} \quad & (\mathbf{s},\mathbf{f}) \in \mathcal{D}.\\
\end{aligned}
\end{equation}

The problem in \eqref{s,f_up} is clearly a convex optimization one that can be solved by various methods including sub-gradient projection \cite{5}, interior point and ellipsoid methods \cite{6,7}. \looseness=-1
\subsection{($\mathbf{w}$, $\mathbf{z})$ update}
From the Lagrangian expression in \eqref{Lagrangian}, the $\mathbf{w}$ update can be found by solving;
\begin{equation}  
\begin{aligned}
\mathbf{w}^{\{l+1\}}&=\argmin_{\mathbf{w}} \|\mathbf{w}-(\mathbf{s}^{\{l+1\}}-\frac{\boldsymbol\lambda_{1}^{\{l\}}}{\rho})\|_{2}^{2}\\
&=\mathbf{s}^{\{l+1\}}-\frac{\boldsymbol\lambda_{1}^{\{l\}}}{\rho},
\end{aligned}
\end{equation}
while that of $\mathbf{z}$ is given by;
\begin{equation}  
\begin{aligned}
\mathbf{z}^{\{l+1\}}&\!=\!\argmin_{\mathbf{z}}\mathbf{1}^{\top}\mathbf{z}\!+\!\boldsymbol\theta^{\{l\}^{\top}}\!(\mathbf{t}^{\{l+1\}}\!-\!\mathbf{z})\!+\!\frac{\rho}{2}\|\mathbf{t}^{\{l+1\}}\!-\!\mathbf{z}\|_{2}^{2}\\
&\!=\!\mathbf{t}^{\{l+1\}}\!+\!\frac{\boldsymbol\theta^{\{l\}}\!-\!\mathbf{1}}{\rho}.
\end{aligned}
\end{equation}
The steps of the ADMM algorithm described in the previous sections can then be summarized as in algorithm \ref{alg:pseudoApriori}.  
\begin{algorithm}[t]
\caption{ADMM algorithm}
\label{alg:pseudoApriori}
\renewcommand{\thealgorithm}{}
\floatname{algorithm}{}
\begin{algorithmic}[1]
    \State Initialize: $\mathbf{w}$, $\mathbf{z}$, $\mathbf{s}$, $\mathbf{f}$, $\boldsymbol\lambda_{1}$, $\boldsymbol\lambda_{2}$, $\boldsymbol\theta$, $\rho$, $k=0$, $v=1$, $u=2$.
    \Repeat
    \For{\texttt{$i \in [n]$}}
        \State \texttt{solve} $a_{i}^{2v}+\frac{u}{v}\left(a_{i}^{2u}-\tilde{t}_{i}a_{i}^{u}\right)-\tilde{x}_{i}a_{i}^{v}=0$
        \State $(d_{i}^{\{l+1\}},t_{i}^{\{l+1\}})=(a_{i}^{*^{v}},a_{i}^{*^{u}})$
      \EndFor
    \State $\mathbf{\hat{d}}=\mathbf{d}^{\{l+1\}}\!+\!\frac{\boldsymbol\lambda_{2}^{\{l\}}}{\rho}$, $\mathbf{\hat{w}}=\mathbf{w}^{\{l\}}\!+\!\frac{\boldsymbol\lambda_{1}^{\{l\}}}{\rho}$ 
    \State $(\mathbf{s}^{\{l+1\}},\mathbf{f}^{\{l+1\}})\!=\!\underset{\mathbf{s},\mathbf{f} \in \mathcal{D}}\argmin\|\mathbf{s}-\mathbf{\hat{w}}\|_{2}^{2}+\|\mathbf{f}-\mathbf{\hat{d}}\|_{2}^{2}$
    \State $\mathbf{w}^{\{l+1\}}=\mathbf{s}^{\{l+1\}}-\frac{\boldsymbol\lambda_{1}^{\{l\}}}{\rho}$
    \State $\mathbf{z}^{\{l+1\}}=\mathbf{t}^{\{l+1\}}+\frac{\boldsymbol\theta^{\{l\}}-\mathbf{1}}{\rho}$
    \State $\boldsymbol\lambda_{1}^{\{l+1\}}=\boldsymbol\lambda_{1}^{\{l\}}+\rho(\mathbf{w}^{\{l+1\}}-\mathbf{s}^{\{l+1\}})$  
    \State $\boldsymbol\lambda_{2}^{\{l+1\}}=\boldsymbol\lambda_{2}^{\{l\}}+\rho(\mathbf{d}^{\{l+1\}}-\mathbf{f}^{\{l+1\}})$ 
    \State $\boldsymbol\theta_{1}^{\{l+1\}}=\boldsymbol\theta_{1}^{\{l\}}+\rho(\mathbf{t}^{\{l+1\}}-\mathbf{z}^{\{l+1\}})$
    \State $l=l+1$
    \Until{convergence}
\end{algorithmic}\end{algorithm}

\section{Numerical results} \label{numerical_results}
In this section, we validate the ability of algorithm \ref{alg:pseudoApriori} in solving problem \eqref{opt3}. For comparison purposes, we use a convex relaxation of \eqref{opt2}, using the $\ell_{1}$ norm in the objective, as a baseline. We did not include any other solution methods discussed in the literature due to the lack of their ability to handle the stability of the system when data fragmentation takes place. Our numerical results consists mainly of two parts; 1) System identification with random data. 2) Identification with real data from a flexible robot arm. In the next parts, we assume that $p=0.5$, i.e., $\ell_{0.5}$. With this selection of $p$, the algorithm converges more quickly and the polynomial root finding problem in \eqref{polynomial} is easier to solve. Numerical experiments were carried out for various values of $p$, i.e., $p\in\{\frac{1}{3},\frac{1}{4}\}$, however, they were not found to outperform the $\ell_{0.5}$ case. Therefore, they are not included in the numerical results section and still under investigation.
\subsection{System identification with random data} \label{sys_iden_random_data}
We consider four data chunks, $T=4$, with 50 samples per chunk, where the samples of each chunk are drawn independently from a symmetric uniform distribution on the interval $\mathcal{I}_{5}$. Input chunks are applied to a randomly generated stable LTI system with a known order, where the initial conditions of the zero input response for each chunk are initialized through samples of zero mean Gaussian distribution with standard deviation $\sigma=10^{-2}$. We assume a uniform gridding of the unit circle into $n=146$ points. As mentioned before in section \ref{proposed_solution}, the denser the grid of the the unit circle is, the better the system is represented but the more complex it will be. From \cite{20}, our choice is a good approximation. Realization noise is added to the LTI system's output, where samples of the noise, $n(k)$, are drawn independently from a uniform distribution on the interval $\mathcal{I}_{0.25}$. We assume a symmetric $2^{m}$-levels, $m=3$, uniform quantizer that maps the entire domain $\mathcal{I}_{\infty}$ to $2^{3}$ levels equally spaced on the interval $\mathcal{I}_{3}$ with quantization step $\Delta=0.8571$. 5 samples per chunk, (10$\%$) of the chunk size, are missing from the quantizer output, where the instances of the missing chunks are random and independent from each other. It is important to highlight that the chunks' sizes and number of missing samples per chunk could be arbitrary and different among chunks, however, we only assumed that these quantities are equal among chunks to simplify the implementation. 

All the other parameters in step 1 of algorithm \ref{alg:pseudoApriori} are initialized through samples from a Gaussian distribution of zero mean and $10^{-1}$ standard deviation. The value of $\rho$ is set to 20. We define a threshold $\bar{\epsilon}$ as the value below which a vector entry is considered zero. The value of the threshold $\bar{\epsilon}$ is chosen such that it is less than $0.5\%$ of the maximum value of the optimal vector $\mathbf{d}$, which makes $\bar{\epsilon}=10^{-3}$ a good choice. The algorithm stops if either $\left\lVert \mathbf{d}-\mathbf{f}\right\rVert_{2}\leq 10^{-2}$ or an iteration budget of 100 iterations is consumed. This budget value was determined through a process of trial and error across several repetitions of the experiment. In some cases, the algorithm's output of $\left\lVert \mathbf{d}-\mathbf{f}\right\rVert_{2}$ converges to a value that is only slightly greater than $10^{-2}$, but very close to it. Figure \ref{convergence} shows the convergence of $\frac{\left\lVert \mathbf{d}-\mathbf{f}\right\rVert_{2}}{\left\lVert\mathbf{f}\right\rVert_{2}}$ with respect to the iteration number for a single run. It can be realized that a budget of around 80 iterations is enough for the algorithm to converge.
We perform two different experiments: 1) A single system is considered and different properties from $\ell_{1}$ and $\ell_{0.5}$ relaxations are compared. 2) Multiple systems with same original order are generated and the different statistical properties are studied.
\subsubsection{Single system experiment} \label{single_system}
In this subsection, we consider the experiment where an input is applied to a stable randomly generated system of order 10. Noise is then added to the output and then applied to the quantizer. The noise values and quantizer setup are as discussed above. Given the sensor outputs, the problem is solved via $\ell_{1}$ and $\ell_{0.5}$ relaxations and the detected system orders and outputs are compared.

\begin{figure}[t]
     \centering
     \begin{subfigure}[b]{0.45\textwidth}
         \centering
         \includegraphics[scale=0.2]{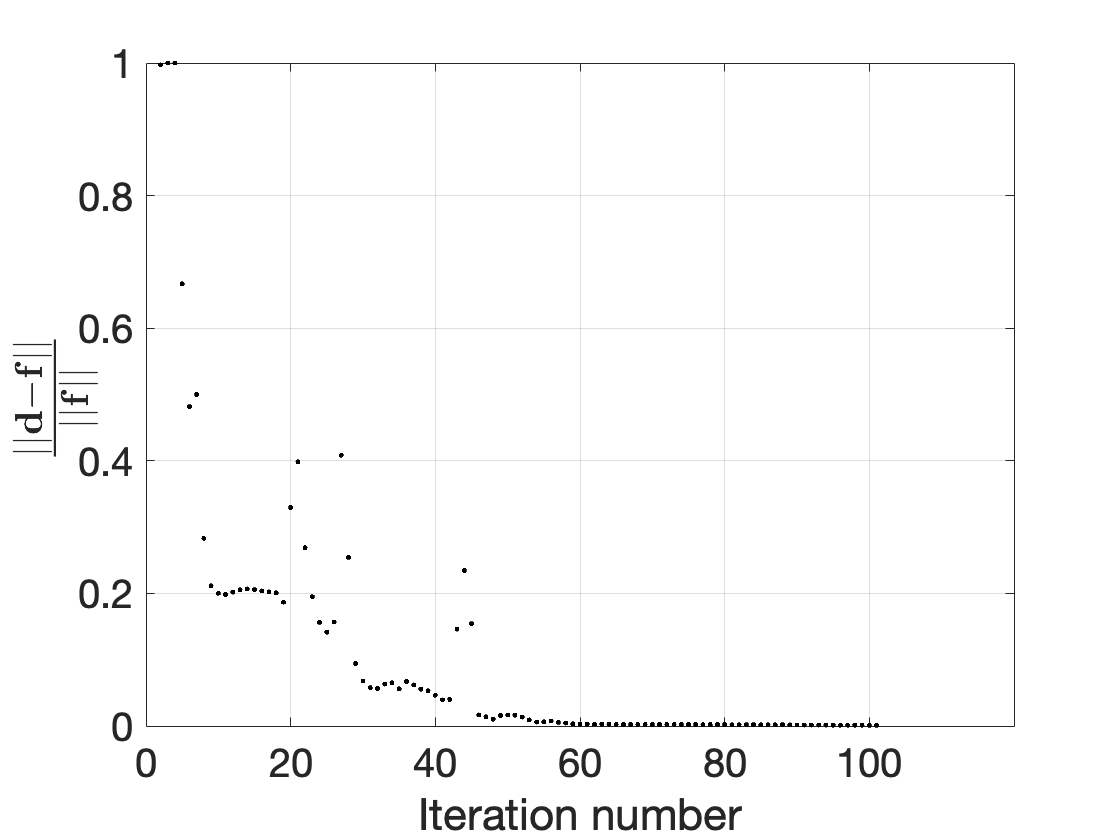}
         \caption{Convergence vs iteration number.}
         \label{convergence}
     \end{subfigure}
     %\hfill
     \hspace{0.25cm}
     \begin{subfigure}[b]{0.45\textwidth}
         \centering
         \includegraphics[scale=0.2]{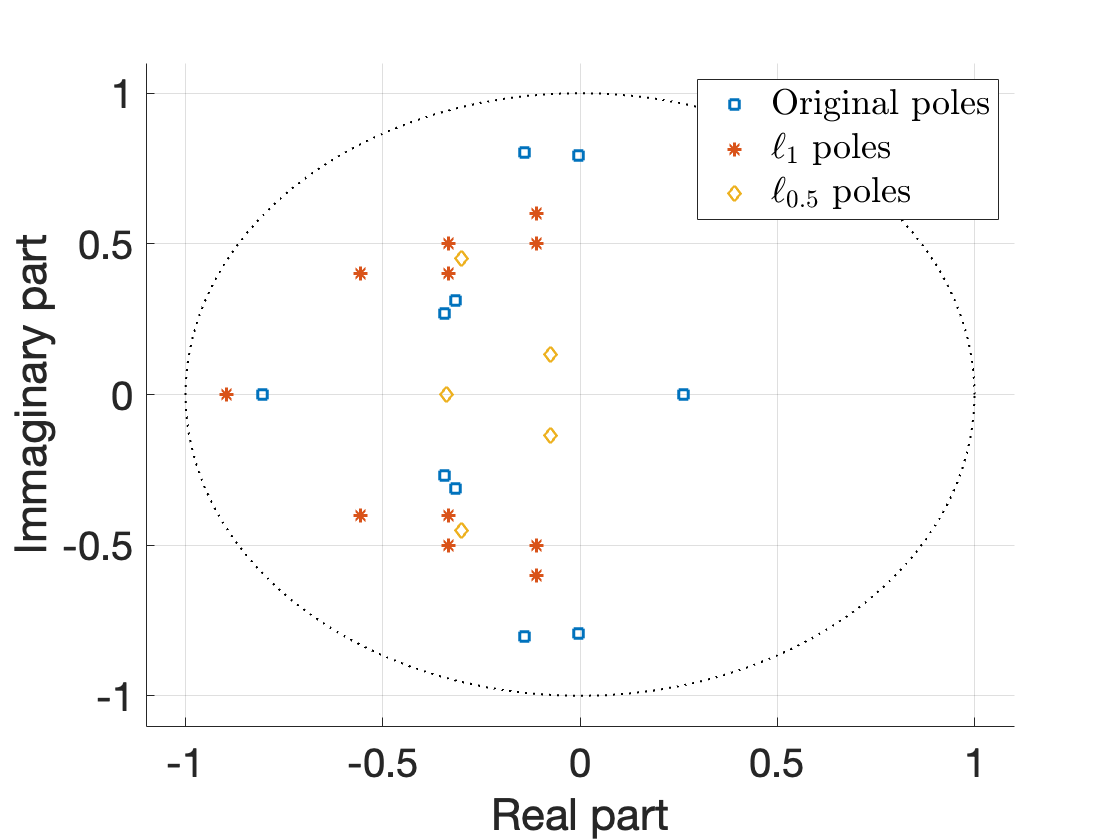}
         \caption{System poles.}
         \label{sys_poles}
     \end{subfigure}
     %\hfill
     \begin{subfigure}[b]{0.45\textwidth}
         \centering
         \includegraphics[scale=0.2]{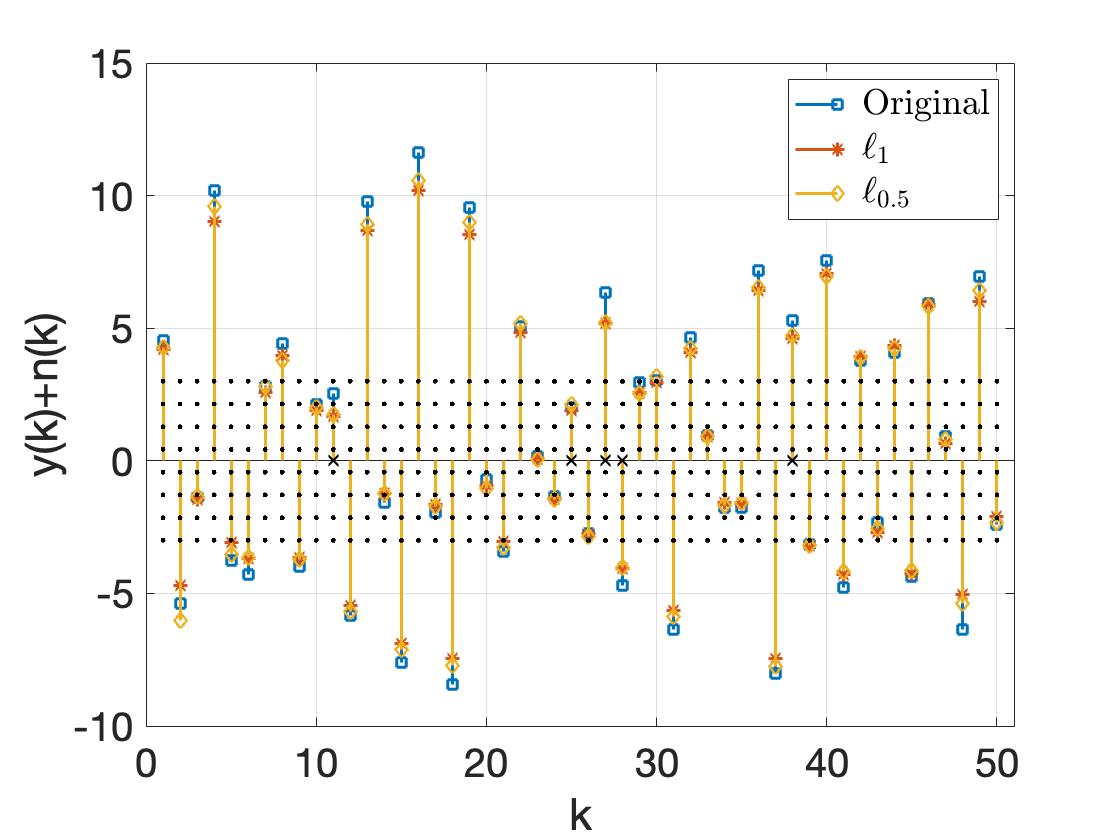}
         \caption{Sensor input}
         \label{sensor_in}
     \end{subfigure}
     %\hfill
     \hspace{0.25cm}
     \begin{subfigure}[b]{0.45\textwidth}
         \centering
         \includegraphics[scale=0.2]{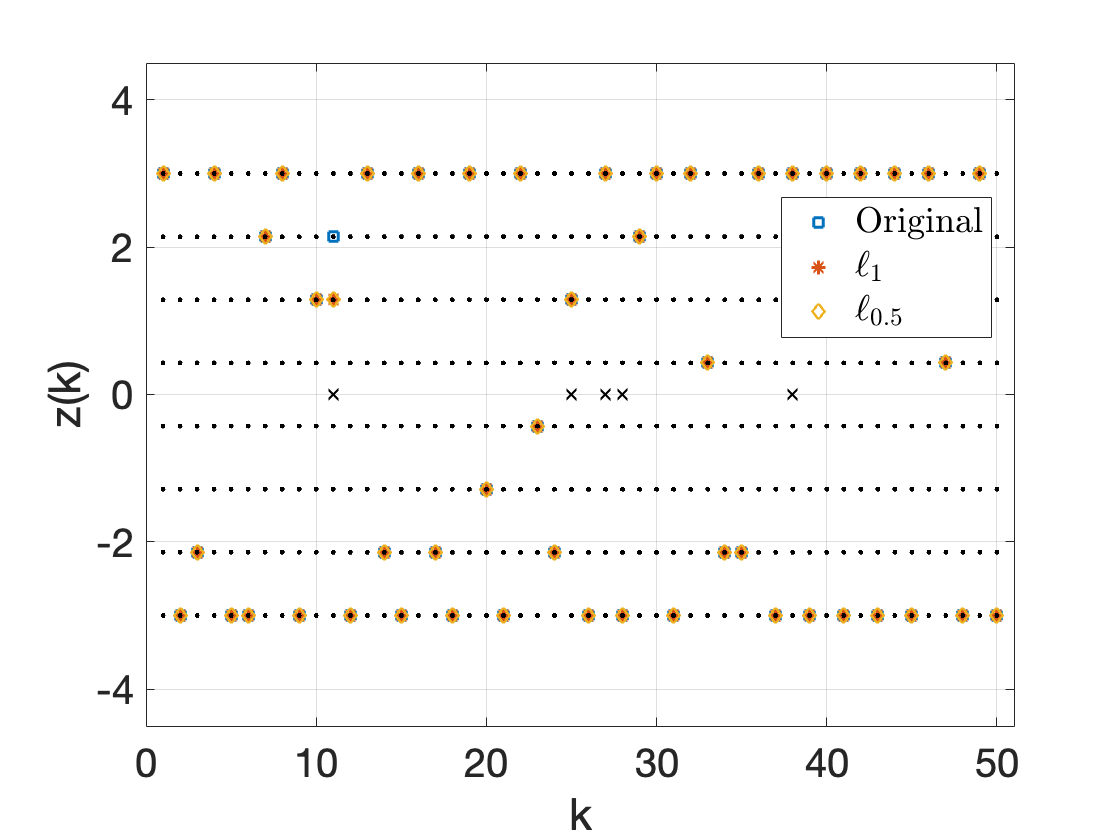}
         \caption{Sensor output.}
         \label{sensor_out}
     \end{subfigure}
        \caption{a) The algorithm convergence. b-d) Single system experiment results. Blue squares, red stars and green diamonds are the original, $\ell_{1}$ and $\ell_{0.5}$ relaxations respectively. Dotted lines in \ref{sensor_in} and \ref{sensor_out} are the used sensor levels. `$\mathbf{x}$' indicates that output data is missing at that instance}
        \label{sing_sys}
\end{figure}

Figure \ref{sys_poles} plots the original system poles vs those that are associated with the non zero coefficients in the vector $\mathbf{a}$ and matrix $\mathbf{B}$ from the $\ell_{1}$ and $\ell_{0.5}$ relaxations' solutions. From the figure, it can be concluded that the $\ell_{0.5}$ detected a system of order 5 which is less complex than the system of order 11 detected by the $\ell_{1}$ relaxation. This outlines the out-performance of the $\ell_{0.5}$ quasi-norm when compared to the $\ell_{1}$ convex relaxation. 

In figures \ref{sensor_in} and \ref{sensor_out}, we plot the sensor input $\hat{y}(k)=y(k)+n(k)$ and output $\mathbf{z}(k)$, vs a finite time horizon $N$ for the fourth chunk. The figures show how accurate the considered relaxations, whether $\ell_{1}$ or $\ell_{0.5}$, can represent the sensor inputs and outputs. We define the sensor input representation error across a time horizon of length $N$ as, $\zeta_{x}^{\text{in}}, x\in\{\ell_{1},\ell_{0.5}\}$ where;
\begin{equation} \label{sensor_in_error}
    \zeta_{x}^{\text{in}}=\sqrt{\sum_{k=0}^{N-1}(\hat{y}(k)-\hat{y}_{x}(k))^{2}},
\end{equation}
with $\hat{y}(k)$ as the noisy output from the original system. For the $\ell_{0.5}$ relaxation, the representation error $\zeta_{\ell_{0.5}}^{\text{in}}$ was found to be equal 3.3204 which is less than that of the $\ell_{1}$ convex relaxation that had a value $\zeta_{\ell_{1}}^{\text{in}}=4.2520$. It is important to note that we are not interested in perfectly fitting the original system's output. However, we aim to fit the sensor's realizations. Hence, we similarly define the sensor output representation error $\zeta_{x}^{\text{out}}, x\in\{\ell_{1},\ell_{0.5}\}$, such that,
\begin{equation} \label{sensor_error}
    \zeta_{x}^{\text{out}}=\sqrt{\sum_{k\in \mathcal{K}_{4}}(z(k)-z_{x}(k))^{2}},
\end{equation}
where, $z(k)$ and $z_{x}(k)$ in \eqref{sensor_error} are the discrete outputs from the original sensor and the considered algorithms while $\mathcal{K}_{4}$ is the set of time indices where the data is available for the fourth chunk. From figure \ref{sensor_out}, it can be realized that $\zeta_{\ell_{0.5}}^{\text{out}}=\zeta_{\ell_{1}}^{\text{out}}=0$. In both figures \ref{sensor_in} and \ref{sensor_out}, the sensor levels are indicated by the dotted horizontal lines. The missing instances are marked by `$\mathbf{x}$' symbol. It can be realized that both algorithms perform a decent job in reconstructing the sensor input and output samples at those missing instances.

% \begin{figure}
% \centering
% \begin{minipage}{.45\textwidth}
%   \centering
%   \includegraphics[scale=0.4]{sys_poles.png}
%   \captionof{figure}{System poles. Blue circles, red stars and green diamonds are the original, $\ell_{1}$ and $\ell_{0.5}$ relaxation poles respectively.}
%   \label{sys_poles}
% \end{minipage}%
% \hspace{0.5cm}
% \begin{minipage}{.45\textwidth}
%   \centering
%   \includegraphics[scale=0.4]{sensor_in_out_ex.png}
%   \captionof{figure}{Sensor input and output. The dotted lines indicate the used sensor levels. The `$\mathbf{x}$' symbol indicates that output data is missing at that instance}
%   \label{sensor_in_out}
% \end{minipage}
% \end{figure}

% \begin{figure}[t]
% \begin{center}
% \includegraphics[scale=.4]{sys_poles.png}
% \caption{System poles.} \label{sys_poles}
% \end{center}
% \end{figure}
% \begin{figure}[t]
% \begin{center}
% \includegraphics[scale=.4]{sensor_in_out_ex.png}
% \caption{Sensor input and output. The dotted lines indicate the used sensor levels. The `$\mathbf{x}$' symbol indicates that output data is missing at that instance} \label{sensor_in_out}
% \end{center}
% \end{figure}
\subsection{Multiple system experiment}
Since the systems that we generate to validate our solution method are random, the main idea in this part is to study the statistical properties of the derived algorithm solution. We perform an experiment where for a given original order, 50 random systems are generated. For each system, the same input is applied and the identification problem in \eqref{opt3} is solved, using the $\ell_{1}$ norm and $\ell_{0.5}$ quasi-norm relaxations, given the quantized realizations from the sensor output.

\begin{figure}[t]
\begin{center}
\includegraphics[scale=0.25]{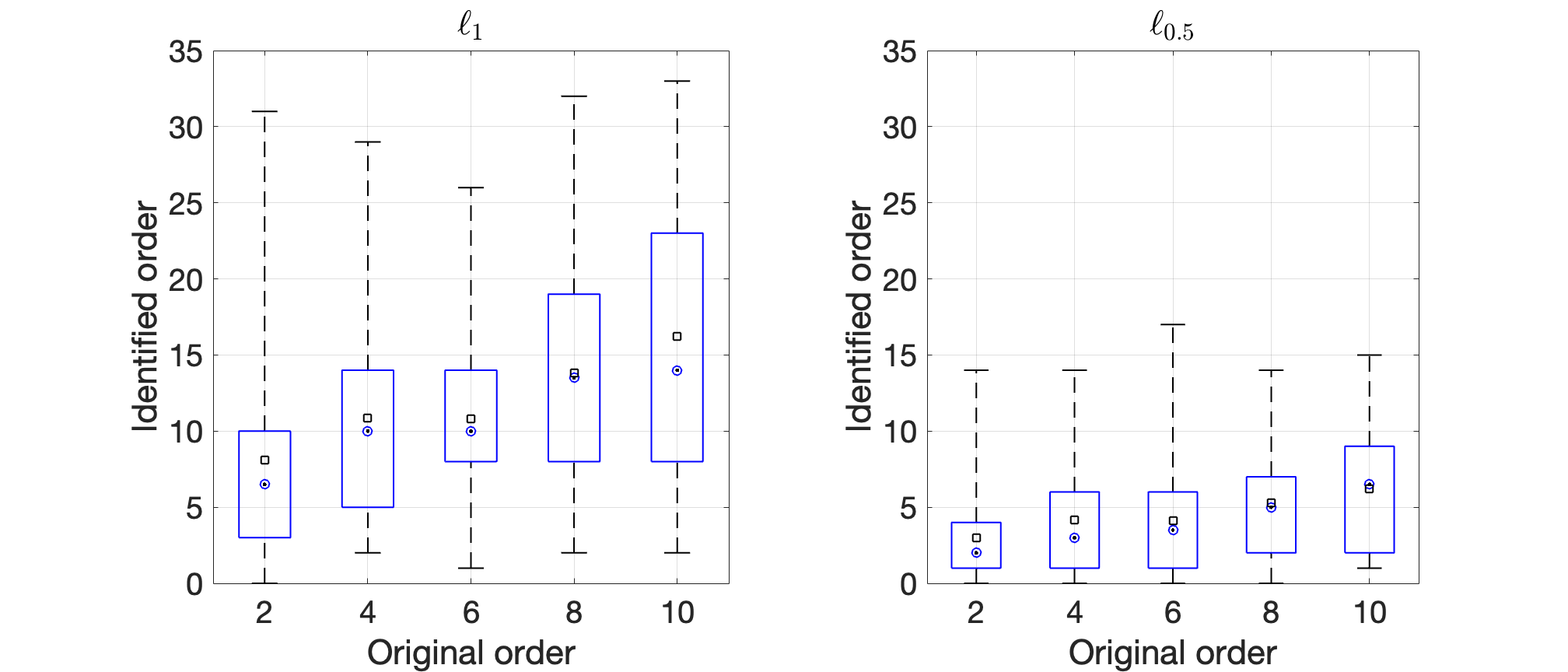}
\caption{Box plot for the system order statistics. Circles with dots and black squares indicate the median and mean values respectively. Bottom/top edges of the boxes are the 25th/75th quantile. The whiskers extend from the minimum (downwards) to the maximum (upwards) value.} \label{avg_box_plot}
\end{center}
\end{figure}

Figure \ref{avg_box_plot} outlines the different statistical properties from the $\ell_{1}$ and $\ell_{0.5}$ relaxations. It can be realized that for all original system orders, the $\ell_{0.5}$ relaxation solution enjoys less mean and median values than its counterpart, i.e., $\ell_{1}$ relaxation. Moreover, the $\ell_{0.5}$ relaxation has a maximum value for each original order that is less than that of the $\ell_{1}$. It can also be realized that in either cases, some systems have a detected order of zero, i.e., the minimum value of the whisker is zero, which means that the estimation of the constant $r$ in \eqref{impulse_res} is enough to describe the I/O relationship. Finally, some systems are detected with higher order than the original, this because the $\ell_{0.5}$ minimization is a non convex problem and hence algorithm \ref{alg:pseudoApriori} converges to a local minimum. Moreover, it motivates that the unit circle should be gridded into more points to increase precision, i.e., $n>146$ mentioned in \ref{sys_iden_random_data}, in expense of computational complexity.

\subsection{System identification using data from a flexible robot arm}
In this part, we consider the identification problem using data collected from the motion of a flexible robotic arm. As described in \cite{daisy_dataset}, the arm is installed on an electrical motor, where, the input represents the reaction torque of the structure to the ground while the output is the acceleration of the arm. The data is composed of 1024 samples, which we divide into 20 chunks of 50 samples each and hence, we drop the last 24 samples of the data set. We assume a uniform $2^{m}$-levels, $m=2$, quantizer that maps $\mathcal{I}_{\infty}$ to $2^{2}$ levels equally spaced on $\mathcal{I}_{0.7}$ with a quantization step $\Delta=7/30$. Similar to as described in \ref{sys_iden_random_data}, we drop $10\%$ of the chunk's samples, where the location of the missing samples are chosen at random. For the $\ell_{0.5}$ quasi norm algorithm, we use the same algorithm initialization as in the previous section while setting $\rho$ to 50 and making the algorithm terminates if a budget of 100 iterations is consumed. 

\begin{figure}[t]
     \centering
     \begin{subfigure}[b]{0.45\textwidth}
         \centering
         \includegraphics[scale=0.2]{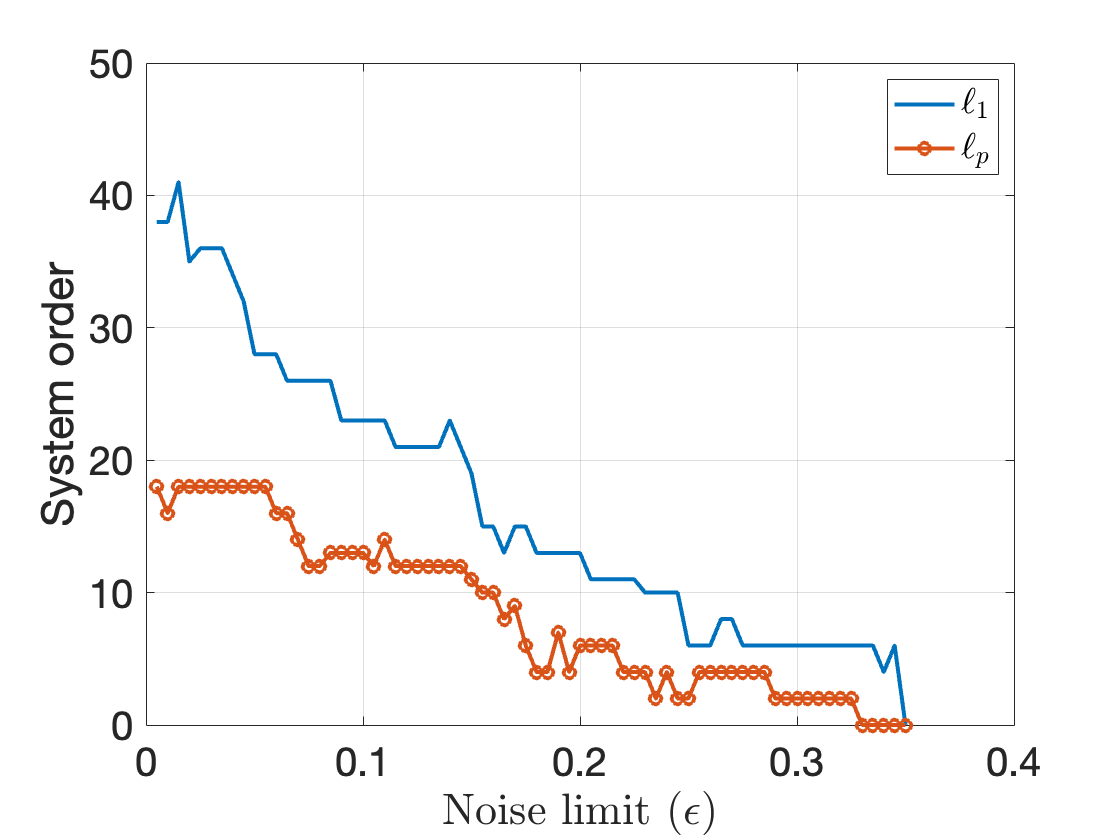}
         \caption{System order vs noise range $\mathcal{I}_{\epsilon}$.}
         \label{order_vs_noise}
     \end{subfigure}
     %\hfill
     \hspace{0.25cm}
     \begin{subfigure}[b]{0.45\textwidth}
         \centering
         \includegraphics[scale=0.2]{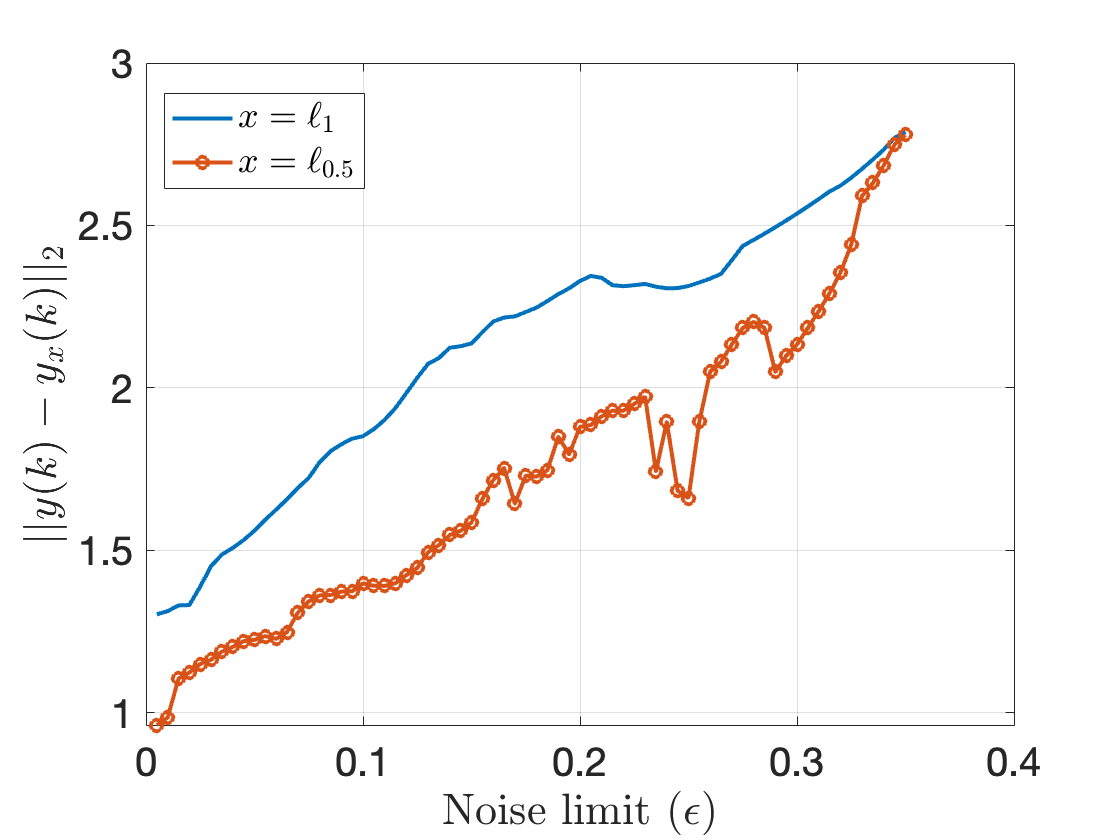}
         \caption{System output error vs noise range $\mathcal{I}_{\epsilon}$.}
         \label{error_vs_noise}
     \end{subfigure}
     \caption{Robotic arm experiment results.}
        \label{rob_arm}
\end{figure}

We report the results for the first available data chunk with a threshold value $\bar{\epsilon}=10^{-3}$. Figure \ref{order_vs_noise} plots the detected system order vs $\epsilon$ which defines the noise boundaries in the range $\mathcal{I}_{\epsilon}$, i.e. $n^{(i)}(k)\in \mathcal{I}_{\epsilon}$. It can be realized from figure \ref{order_vs_noise} that, for both the $\ell_{1}$ norm and the $\ell_{0.5}$ quasi norm, the detected order decreases with the increase of $\epsilon$. This is intuitive because on increasing $\epsilon$, the size of the feasibility set increases which enables systems of lower orders to be explored. Moreover, a momentarily increase in the system order can happen while increasing $\epsilon$. This is because we mainly aim to minimize a relaxed version in \eqref{opt3} instead of the original one in \eqref{opt1} and hence, more non zero low value entries can decrease the objective of \eqref{opt3}. For all values of $\epsilon$, the $\ell_{0.5}$ quasi norm algorithm detects a lower order than the $\ell_{1}$ convex relaxation. For a chunk of size 50 samples, it can be realized that the $\ell_{1}$ norm objective recovers systems of orders $\sim$ 35:40, for small values of $\epsilon$. This indicates that the $\ell_{1}$ relaxation tends to over fit the data for low values of $\epsilon$, while the $\ell_{0.5}$ one aims to recover a model which accurately represents it. The norm of the noiseless system output (quantizer input) error, denoted by $\left\lVert y(k)-y_{x}(k) \right\rVert_{2}, x\in\{\ell_{1},\ell_{0.5}\}$, is plotted in figure \ref{error_vs_noise}. With the same justification as in Figure \ref{order_vs_noise}, more systems that might have a lower order but higher output error are added to the feasible set when the value of $\epsilon$ is increased. We are not concerned in exactly fitting the output of the original system, as was covered in section \ref{single_system}. However, our goal is to choose the system from the feasibility set that fits the realizations of the sensor while having the lowest order.

% \begin{figure}[t]
% \begin{center}
% \includegraphics[scale=0.4]{order_vs_noise.png}
% \caption{System order vs noise range $\mathcal{I}_{\epsilon}$.} \label{order_vs_noise}
% \end{center}
% \end{figure}

\section{Conclusion} \label{conclusion}
In this paper, we presented an approach that aims to find the least order system that is compatible with fragmented quantized realizations. This approach allows for the use of a priori information on the system and fragmented measurements of the output. The algorithm is based on an ADMM approach that aims to solve an $\ell_{p}$ quasi-norm objective by dividing the optimization over the variables through iteratively solving simpler sub-problems. The algorithm is tested on a synthetic data set, that is randomly generated, and a realistic data set collected through the measurement of the movement of a robotic arm. Numerical results presented show that the algorithm is very effective in obtaining low complexity explanations of the data collected. Further effort is being put into analyzing the convergence of the proposed algorithm, improving the numerical performance and its extension to continuous-time systems.

% \section*{Funding}

% An unnumbered section, e.g.\ \verb"\section*{Funding}", may be used for grant details, etc.\ if required and included \emph{in the non-anonymous version} before any Notes or References.

% \section*{Notes on contributor(s)}

% An unnumbered section, e.g.\ \verb"\section*{Notes on contributors}", may be included \emph{in the non-anonymous version} if required. A photograph may be added if requested.

% \section*{Nomenclature/Notation}

% An unnumbered section, e.g.\ \verb"\section*{Nomenclature}" (or \verb"\section*{Notation}"), may be included if required, before any Notes or References.

% \section*{Notes}

% An unnumbered `Notes' section may be included before the References (if using the \verb"endnotes" package, use the command \verb"\theendnotes" where the notes are to appear, instead of creating a \verb"\section*").

\bibliography{Ref}
\bibliographystyle{IEEEtran}

\end{document}